\documentclass[aps,twocolumn,groupedaddress]{revtex4}
\usepackage{graphicx}

\def\be{\begin{equation}}
\def\ee{\end{equation}}
\def\bea{\begin{eqnarray}}
\def\eea{\end{eqnarray}}
\def\bw{\begin{widetext}}
\def\ew{\end{widetext}}

\begin{document}

\title{Morphology of axisymmetric vesicles with encapsulated 
filaments and impurities
}

\author{Akiyoshi {\sc Shibuya}, Yukio {\sc Saito} and Hiroyuki {\sc Hyuga}}
\email[]{ashibuya@phys.keio.ac.jp}
\email[]{yukio@rk.phys.keio.ac.jp}
\email[]{hyuga@rk.phys.keio.ac.jp}
\affiliation{Department of Physics, Keio University, Yokohama 223-8522}

\date{July 17, 2000}

\begin{abstract}
The shape deformation of a three-dimensional axisymmetric 
vesicle with encapsulated filaments or impurities is 
analyzed by integrating a dissipation dynamics. 
This method can incorporate systematically the constraint of a fixed surface
area and/or a fixed volume.
The filament encapsulated in a vesicle is assumed to take a form of
a rod or a ring so as to imitate 
cytoskeletons.
In both cases, results of the shape transition of the vesicle 
are summarized in phase diagrams in the phase space 
of the vesicular volume and a rod length or 
a ring radius. 

We also study the dynamics of a vesicle with impurities coupled 
to the membrane curvature. 
The phase separation and the associated shape deformation
in the early stage of the dynamical evolution
can well be explained by the linear stability analysis.
Long runs of simulation demonstrate the nonlinear coarsening 
of the wavy deformation of the
vesicle in the late stage.
\end{abstract}

\maketitle


\section{Introduction}
\label{sec1}

Amphiphilic molecules in the aqueous solvent assemble into a bilayer 
membrane, and form a closed vesicle. A typical example is an 
artificial lipid-bilayer, which serves as a model for biomembranes. 
Vesicles of a lipid-bilayer membrane have been studied
as a model system of living cells over a couple of decades. 
In particular, 
after Helfrich proposed an elastic continuum model of the membrane 
\cite{helfrich73},
many works have been made about the morphology of the vesicles.
Even in the case of a single component, vesicles take various shapes, 
and their shapes are classified 
as spherical, prolate, oblate, pear, stomatcyte and so on, 
according to Seifert et al.\cite{udo+91} \par

Actually, biomembranes might be more complex by containing many 
components such as several different species of lipids or 
membrane proteins. 
Impurity lipid molecules and proteins can move freely within the 
two-dimensional liquid state of the membrane and perform diffusive motion.
Via the coupling to the membrane curvature, 
impurities may affect the shape of the vesicle. 
Also the vesicle can enclose cytoskeletons in its interior space. 
Cytoskeletons consist of microtubles 
and/or intermediate filaments, and form a three-dimensional network 
inside a cell to support the cell shape. 
Also for the cell locomotion, the elongation and the shortening of 
filamental proteins in the vesicle are required.\cite{bray92}
Thus in the study of 
living cells, the consideration of the interaction between lipid bilayer 
membranes and such encapsulated materials is indispensable. \par

To study the role of such materials, several experiments have been 
made. Hotani and Miyamoto\cite{hotani+90} established a model 
system where microtubules polymerize in a vesicle. A growing 
microtuble pushes the vesicle membrane at contact points and 
deforms it into the shape resembling a Greek letter $\phi$. 
There have been many experiments and theoretical 
analyses on this phenomenon. 
It was established that to realize the 
$\phi$-shape,  the osmotic pressure inside the vesicle has to be 
higher than the outer pressure,\cite{umeda+98,morikawa+99} 
or  the volume of the vesicle has to be fixed.
\cite{fygenson+99,heinrich+99} It is also found that 
the monopolar asymmetric $\phi$-shape has a lower energy 
and is stabler than 
the bipolar mirror symmetric shape. \par

When the filaments in the vesicle are soft such as actin filaments, 
they bend and form a ring in the vesicle. The experiment of such cases 
was performed by Miyata and Hotani\cite{miyata+92}. They observed 
that ring filaments in a vesicle push the vesicle outward 
and deform the spherical vesicle into a disk shape. 

Shape transformation caused by the phase separation within the 
membrane occurs in many important natural phenomena 
such as cell locomotion, fusion, secretion, 
endocytosis, phagocytosis etc. 
Furthermore, various experiments have been performed 
on this type of shape 
changes in multi-component vesicles. 
One example of the experiments is 
on the transition from a biconcave shape of 
erythrocytes to a crenated one (echinocytosis)
\cite{deuticke+68,allan+75}. 
Another one is on a shape deformation 
induced by the phase separation of the amphiphiles composing the 
membrane\cite{gebhardt+77}. Also there is an experiment on a shape 
deformation caused by proteins\cite{bradley+99} 
or polymers which are anchored in membranes\cite{tsafrir+01}. 
In these cases, vesicles transform from the spherical shape 
to the one with a budding or from a tubular shape
to a pearling one. 
The effect of impurity on the 
vesicle morphology was first analyzed 
by S. Leibler\cite{leibler86}. 
He showed that the impurities diffusing within 
the two-dimensional membrane couple to the curvature 
and destabilize the flat membrane. 
Recently, dynamical simulations of 
two-component vesicles have been performed \cite{taniguchi+96} 
and the coalescence process of the budding vesicle was 
obverved\cite{Kumar+01}.

In this paper, we study morphology of a three-dimensional axisymmetric 
vesicle with encapsulated filaments or embedded impurities, 
by solving purely dissipative dynamics. 
In section \ref{sec2},  the model Hamiltonian of a vesicle is 
introduced, and its dissipative dynamics is formulated.
In section \ref{sec3}, the shape of the vesicle 
encapsulating a rod filament is studied. As a linear 
filament elongates, vesicular shape alters from a sphere 
to a rugby-ball, and to a sphere with tubular protrusions. 
The shape transition is summarized in 
a phase diagram in the phase space of the vesicular 
volume  and a filament length. 
In section \ref{sec4}, we study the case of ring filaments. 
When filaments in the vesicle are soft such as actin filaments, 
they bend and form a ring. 
As the filament extends their length, the ring pushes 
out the vesicle membrane and 
deforms the prolate vesicle into the oblate one. 
In the case of cell division, the ring filament shrinks. 
Relations between the ring-force and the 
vesicle shape or the ring radius are studied in this case, and 
the morphological transition is summarized in the phase diagram.
In section \ref{sec5}, 
we study the  effect of impurities embedded in the membrane 
on the vesicle shape, and observed that impurities 
diffusing within the vesicle surface induce curvature instability. 
A linear stability analysis around a spherical vesicle 
is confirmed in our simulation 
of the vesicle under several conditions. We have further simulated 
the nonlinear coalescing process in the late stage. 
Whole analysis is summarized in section \ref{sec6}.

\section{Dissipation Dynamics of Membranes}
\label{sec2}

The shape of a membrane vesicle is characterized by two parameters
$s_i, ~(i=1, 2)$ such that the position vector 
$\vec r$ of the membrane surface in three dimensional space 
is represented as $\vec r(s_1,s_2)$.
With this parametrization, the metric tensor is given by 
$g_{ij}=\partial_i \vec r \cdot \partial_j \vec r$, and its
determinant is denoted as $g=det\{g_{ij}\}$. Here $\partial_i$ 
denotes the partial differentiation by 
$s_i$ as $\partial _i= \partial/\partial s_i$.
The surface area element is given by $ dA= \sqrt{g} d^2s $ 
and the volume element by 
$dV= dA (\vec r \cdot \vec n)/3$ where $\vec n$ is the surface
normal vector.\par
The energy of the system consists of the bending elastic energy, 
and the surface 
and the volume contributions as
\be
E= \frac{\kappa}{2} \int (H-C_0)^2 dA +\sigma A + PV .
\label{1}
\ee
Here $\kappa$ is the bending rigidity, 
$H$ the mean curvature defined as the sum of two principal 
curvatures $C_1$ and $C_2$ as $H=C_1+C_2$, 
$C_0$ the spontaneous curvature, $\sigma$ the surface tension, and
$P$ the pressure. In a dissipative dynamics,
the system evolves to reach minimum energy configuration dictated by
the energy (\ref{1}). To achieve this purpose, we assume a purely 
dissipative dynamics with the Rayleigh dissipation function of the form 
\be
F_d = \frac{\eta}{2} \int |\vec v|^2 dA .
\label{2}
\ee
with a viscosity coefficient $\eta$. 
The equation for time evolution is then written as
\cite{marsili+96,langer+92,cantat+00}
\be
\frac{\delta F_d}{\delta \vec v} = - \frac{\delta E}{\delta \vec r} .
\label{3}
\ee
This gives, in general, the shape evolution equation as
\be
\partial_t \vec r =- \frac{1}{\eta \sqrt{g}} \frac{\delta E}{\delta 
\vec r} = v_n \vec n= (v_c -\eta^{-1} \sigma H -\eta^{-1} P) \vec n ,
\label{4}
\ee
where the normal component of the velocity due to the curvature 
elasticity is given by
\be
\eta v_c  
= \frac{\kappa}{2} \{ H(C_1-C_2)^2 + 2\tilde \Delta H + 4C_1C_2C_0 \} .
\label{5}
\ee
Here $\tilde \Delta = g^{-1/2} \partial_i ( g^{1/2} \partial^i)$ 
is the Beltrami-Laplace operator. 
This evolution equation (4) is explicitly derived by 
Marsili et al.\cite{marsili+96} in section III.A.1-3 in their paper.
Shape evolution generally induces variations of the surface area or 
the volume enclosed by the vesicle as\cite{marsili+96,langer+92,cantat+00}
\bea
\frac{d A}{d t} 
&= \int  H (v_c -\eta^{-1} \sigma H -\eta^{-1} P) dA,
\label{6a} \\
\frac{d V}{d t} &= \int  (v_c -\eta^{-1} \sigma H -\eta^{-1} P) dA .
\label{6b}
\eea
If the area of the vesicle is to be fixed, the tension $\sigma$ should be 
regarded as a Lagrange multiplier, and has to be adjusted so as to 
satisfy $dA/dt=0$. 
If the volume as well as the surface area are to be fixed, 
both the tension $\sigma$ and the pressure $P$ 
have to be adjusted to yield: $dA/dt= dV/dt =0$. 

For an axi-symmetric vesicle, the position vector is conveniently 
written in the cylindrical coordinate as $(\rho, \varphi, z)$, where 
the $z$ axis is chosen to be the axis of rotational symmetry. 
As for the two parameters, we choose an arclength 
$s_1= s= \int \sqrt{(dz)^2+(d \rho)^2}$ and the angle $s_2= \varphi$. 
The arclength is measured from the bottom of the vesicle 
on the symmetry $z$ axis. 
The rotational symmetry tells us that the shape is independent of the 
second parameter $s_2=\varphi$. Then the metric is 
$\sqrt{g}=\rho$, the angle-integrated area element 
is $dA= 2\pi \rho ds$ and the volume element is 
$dV= \pi \rho^2 z_s ds$. 
Since the shape is independent of the rotation angle 
$\varphi$, we consider hereafter a section at $\varphi=0$, and the 
vesicle contour is given in $(\rho,z)$-space. 
The normal and tangential vectors in this space are given as 
\be
 \vec n= (z_s, - \rho_s), \quad \vec t= (\rho_s, z_s),
\label{7}
\ee
and
two principal curvatures are calculated to be 
\be
C_1=  z_s/\rho, \quad C_2= \rho_s z_{ss} - z_s \rho_{ss} .
\label{8}
\ee
Here the subscript $s$ denotes differentiation by 
$s$, for instance, $z_s=\partial_s z$. 
The Beltrami-Laplace operator is represented as 
\be
\tilde \Delta = \frac{1}{\rho} \partial_s
\left( \rho \partial_s \right) .
\label{9}
\ee
The variation of the elastic energy produces only the normal component of the 
velocity, because only this component has the physical significance.
Nevertheless, we introduce a tangential component of the velocity, 
which corresponds to the rearrangement of the parametrization $s$, 
or a {\it gauge}, on a vesicle surface. 
In the practice of shape simulation, we deal with only a discrete set 
of points (called grid points, hereafter) on the membrane. 
Even if these grid points are initially 
prepared in equidistance, 
they will dilatate or contract locally during the shape evolution. 
In order to redistribute grid points in equal separation, 
we must add an appropriate tangential velocity $v_t(s) \vec t $ at a 
grid point of an arclength $s$. 
Since the total arclength $L$ varies in time, 
the condition of equal separation of grid points 
are expressed by imposing the invariance of a
relative arclength, namely $d(s/L)/dt=0$. By this condition,  
we get the tangential velocity\cite{langer+92,cantat+00} 
\be
v_t(s)= -\int_0^s v_nC_2 ds + \frac{s}{L} \int _0^L v_n C_2 ds .
\label{10}
\ee
The evolution of a vesicle is then obtained by numerically 
integrating the equation 
\be
\frac{\partial \vec r}{\partial t} = v_n \vec n + v_t \vec t .
\label{11}
\ee

\section{Encapsulation of Rod Filament}
\label{sec3}

We consider here vesicle deformation induced by the elongation 
of an encapsulated filaments such as microtubles.
The filaments are rigid and extend straight in the vesicle.
They push and deform the incarcerating vesicle.
There have been several studies on this problem so far.
\cite{umeda+98,morikawa+99}
We first reproduce known results by the present method. 
Since the encapsulated filament is assumed to remain straight, 
it exerts pressing force at both end points 
or poles at $s=0$ and $s=L$ as $F \delta_0(s)$ and $F \delta_0(s-L)$,
respectively. 
Here the invariant form of 
the delta function in the parameter space is given 
\cite{marsili+96} as 
\be
\delta_0(\vec s-\vec s')=\frac{\delta(\vec s-\vec s')}{\sqrt{g}} .
\label{12}
\ee
In the equilibrium case, the forces balance equation around the pole
is given by  
$\eta v_n(s)+ F \delta_0(s)=0$. 
By integrating over a very small area around the pole $s=0$ with a 
radius $\rho$, 
the second term gives simply $F$, 
whereas in the first term only the 
 Beltrami-Laplace term gives rise to nonvanishing contribution as 
\be
\kappa \int \rho^{-1} \partial_s (\rho \partial_s H) 2 \pi \rho ds 
\approx 2 \pi \kappa \rho \partial_{\rho} H .
\label{13}
\ee
Here we use the approximation such that around the pole at $s=0$, 
the profile is almost horizontal and $ds=d \rho$. 
The integration of the force balance equation then yields the curvature 
\cite{umeda+98} as 
\be
H(\rho)= H_0- \frac{F}{2 \pi \kappa} \log \rho/\rho_0 ,
\label{14}
\ee
where $H_0$ and $\rho_0$ are appropriate constants.
This expression implies that the curvature diverges at 
the end $s=0$ ($\rho=0$) and the same type of divergence 
appears at the other end ($s=L$).
Even though this diverging curvature, gives rise to 
only finite contribution to 
the bending energy, the vesicle shape at the poles 
cannot be derived by the dynamics governed by the curvature alone. 
To overcome this difficulty, 
both poles are pushed outwardly until the 
pole-separation $\ell$ reaches the prescribed length of the rod 
filament. 
When the pole-separation exceeds the prescribed distance $\ell$, the
poles are set free from external force and the distance decreases due 
to the curvature force by the nearby surface. 
The force applied on the poles are estimated from 
Eq.(\ref{14}) by the curvature differences between $\rho_1$ and $\rho_2$ 
close to the poles as 

\be
F= 2 \pi \kappa \frac{H(\rho_2)-H(\rho_1)}{\log \rho_1/\rho_2} .
\label{15}
\ee

Simulation results are normalized in the dimensionless form. 
The length is measured in unit of $R_0$, which is the radius of the sphere 
with a surface area of the vesicle, and the energy by $8 \pi \kappa$. 
Then the volume is normalized
by $4\pi R_0^3/3$, 
the pressure $P$ as $p=P R_0^3/\kappa$ and
the force $F$  as  $f=F R_0/4 \pi \kappa$, respectively. 
In the simultaion the spontaneous curvature $C_0$ 
is  set to be 0.

\begin{figure}[h]
\begin{center} 
\includegraphics[width=7cm]{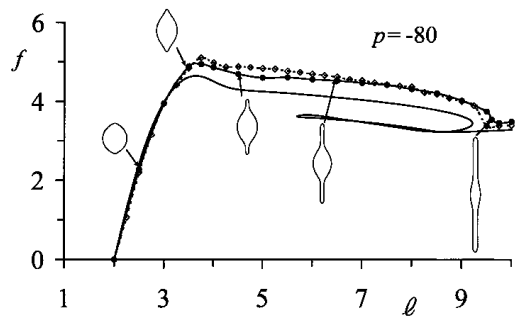}
\end{center} 
\caption{
Separation distance between two poles $\ell$ versus the strength of
the force $f$ for a vesicle 
under a pressure $p=-80$. Data obtained in the present analysis 
are depicted by the symbol $\bullet$. 
A solid curve represents data obtained by the variational 
method by Umeda {\it et al.} \cite{umeda+98}. Results obtained by 
the Monte Carlo simulation \cite{morikawa+99} are represented by 
the symbol 
$\diamond$.  }
\label{fig1}
\end{figure}


We first summerize the results of the simulations with a fixed osmotic 
pressure $p$. By starting from the mirror-symmetric shape of the 
vesicle with symmetric forces at both poles, 
we get the vesicle with a mirror symmetry about the equator. 
The shapes as well as the length-force relation are shown 
in Fig.\ref{fig1}, in good agreement with those obtained previously. 
\cite{umeda+98,morikawa+99} 
With a negative osmotic pressure, 
the elongation of the filament length $\ell$ inside a spherical 
 vesicle leads first to the increase of the strength of the force $f$ 
 applied 
to both poles, but eventually $f$ reaches 
the maximum when 
two poles start to protrude. 
The maximum value of $f$ turns out to be a little 
larger than the value obtained by Umeda {\it et al.}, probably 
due to the relatively small number of grid
 points ($N=120$) in our simulation. 
By applying asymmetric forces at both poles in the initial stage of the
simulation, 
we can get an asymmetric shape. This is due to the fact that 
the asymmetric shape has a lower energy and is stabler 
than the mirror symmetric one, as is already discussed previously. 
\cite{umeda+98,morikawa+99}

\begin{figure}
\begin{center} 
\includegraphics[width=5.5cm]{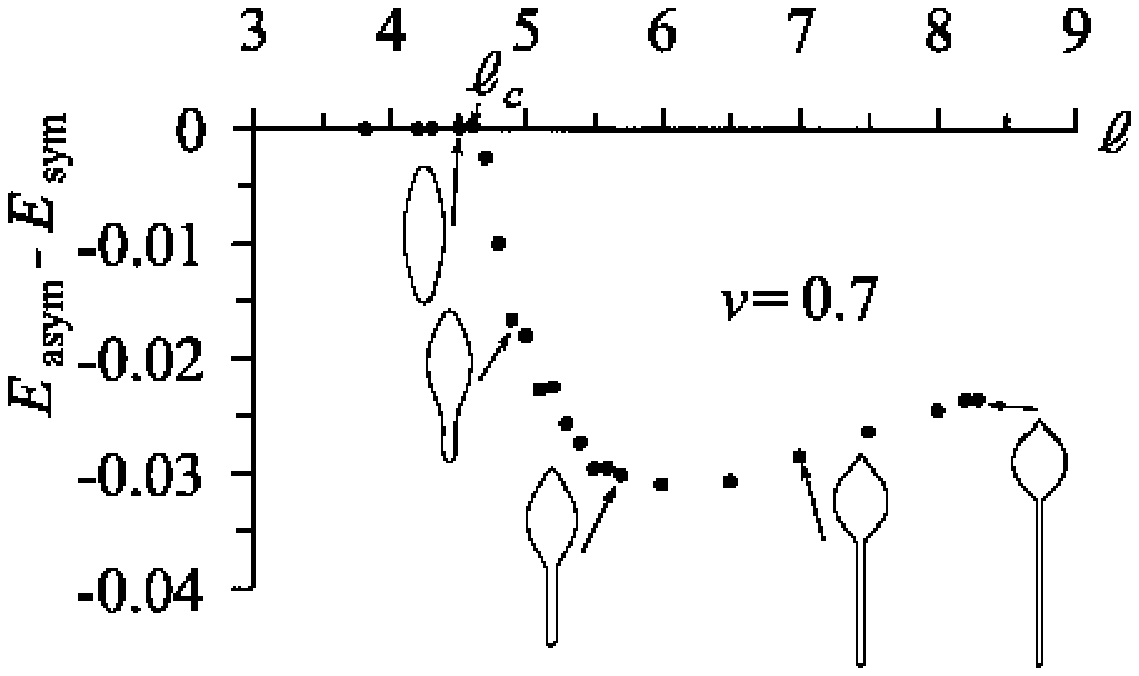}\\
\quad (a)\\
\includegraphics[width=5.5cm]{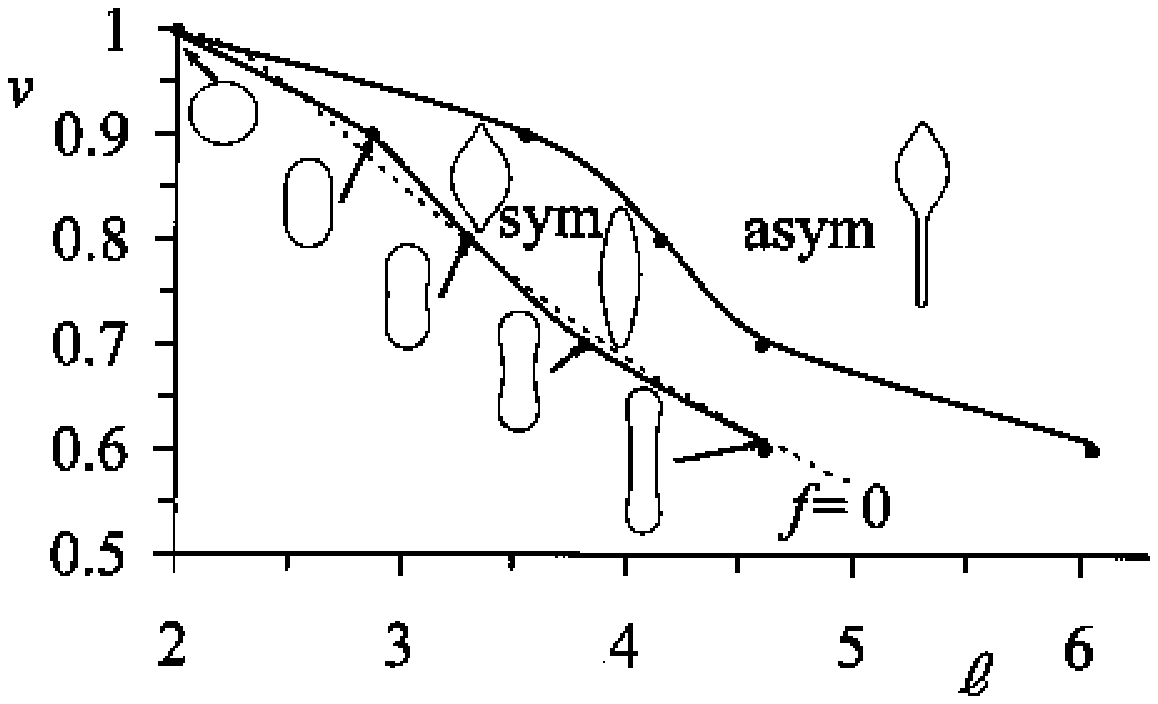}\\
\quad (b)\\
\includegraphics[width=5.5cm]{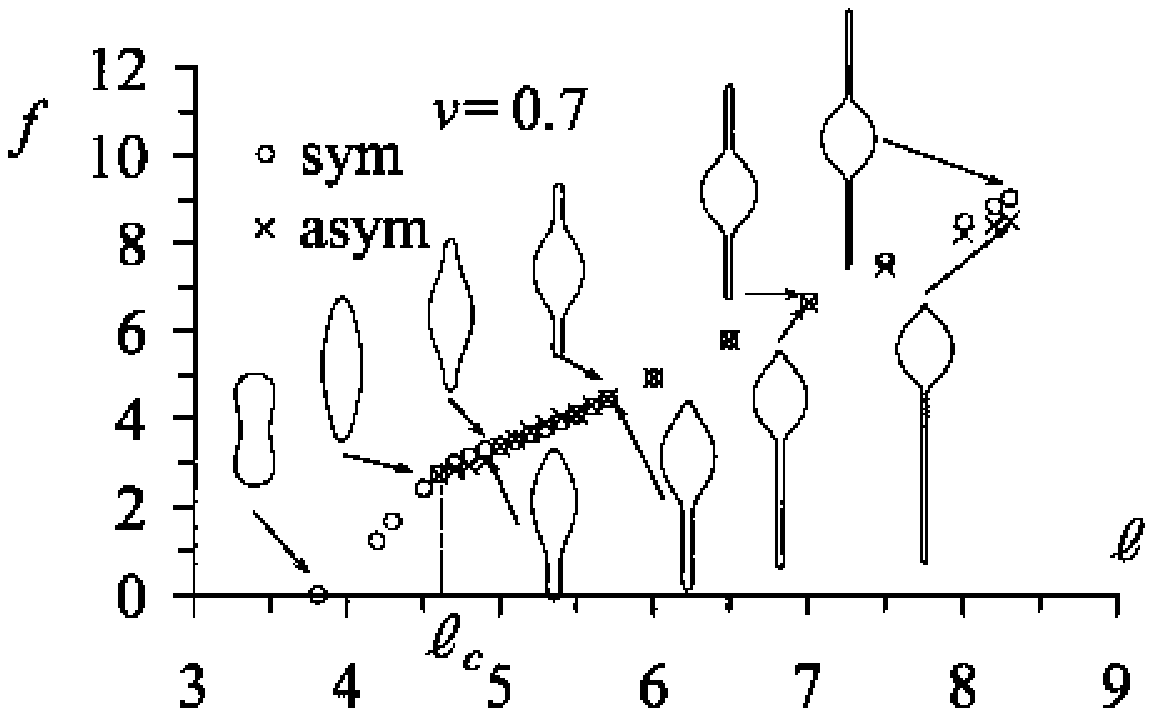}\\
\quad (c)
\end{center} 
\caption{
(a) Energy difference of the assymmetric and the symmetric vesicle 
as a function of the rod length $\ell$. 
The volume of the vesicle is fixed 
at $v=0.7$. 
(b) Morphology phase diagram of the vesicles with various volume 
$v$ and the rod length $\ell$. A dashed curve indicates the 
cylinders with hemishperical caps at both ends. 
(c) The strength of the force $f$ versus the rod length $\ell$ at 
a volume $v=0.7$. The critical length for protrusion is $\ell _c=4.6$.}
\label{fig2}
\end{figure}

One of the advantage of the present method is that
 the shape of a vesicle with a  fixed 
volume can be studied easily and quickly.\cite{heinrich+99} 
When the length of the rod is short, the vesicle takes a 
mirror symmetric shape independent of the initial condition. 
On the other hand when the rod is long and two protrusions appear,
 two distinct shapes, that is symmetric 
and asymmetric shapes, become possible. 
Both shapes have similar energies, but 
the energy difference between these two shapes at a volume $v=0.7$ 
shown in Fig.\ref{fig2}(a)  
indicates that the asymmetric shape is energetically 
favorable than the symmetric one. 
The morphological phase diagram is summarized in the phase space of
the volume $v$ and the rod length $\ell$ in Fig.\ref{fig2}(b).
Some typical shapes are shown therein. 
The phase boundary marked by $f=0$ represents the $v - \ell$ 
relation for the free prolate shape without any force at poles. 
In order to decrease the volume under the condition of a fixed surface area, 
the vesicle changes its shape from sphere to prolate. 
In fact, the $v - \ell$ relation can be well explained by assuming that the
vesicle takes the shape of a cylinder with hemispherical caps at both
ends. The radius of the sphere $r$ and the length of the cylinder $\ell-2r$
should be determined from the conditions of constant surface area 
$4 \pi r^2+ 2 \pi r (\ell- 2r) = 4 \pi$ and the constant volume
$4 \pi r^3/3+ \pi r^2 (\ell-2r) = 4 \pi v/3$.
The $v-\ell$ relation so determined is denoted by a dashed curve 
in Fig.\ref{fig2}(b).
For a cylinder part, one of the curvature vanishes, $C_2=0$.
In the actual situation the vesicle lowers its elastic energy by selecting 
an appropriate value of $C_2$, and thus the phase boundary $f=0$ deviates 
a little from the curve expected for the simplified cylindrical shape.

When the force is applied on two poles of the vesicle with a fixed volume, 
the rod length $\ell$ increases.
For instance at $v=0.7$, the strength of the force $f$ increases as 
$\ell$  
as shown in Fig.\ref{fig2}(c). 
Initially, as the filament elongates, $f$ increases rapidly 
until the poles begins to push out. 
Once one or both poles start to protrude, the increment of $f$ gets 
mild. As the rod gets long, 
the tubular protrusion becomes thin with a small diameter, and the 
central bulb gets more rounded. 

The shape of the vesicle under a strong force might be approximated by the 
form of a cylindrical tube attached to a spherical bulb. 
In order to fulfil the volume conservation, the sphere covers
the volume $4 \pi v/3$ such that its radius is just $v^{1/3}$.
Since the sphere gives only the area $4 \pi v^{2/3}$, the remaining
area is covered by the fine cylindrical tube with an extension
$\ell'$ and the radius 
$r \approx 2(1-v^{2/3})/ \ell'$. 
For $\ell' \rightarrow \infty$, the volume of the cylinder vanishes
$\pi r^2 \ell' \sim 1/ \ell' \rightarrow 0$. Therefore, the $rhs$ region
of the
phase diagram in Fig.2(b) extends to infinity except at $v=1$.

\section{Encapsulation of Ring filament}
\label{sec4}

When the filaments encapsulated are soft like actin filaments, 
they tend to bend and to form a ring in a vesicle. 
As their degree of polymerization or their length increases, 
the ring radius expands. 
On the contrary, in the case of cell division, an actin filament 
ring shrinks. 

The force exerted by the ring at $z=0$ can be 
estimated by assuming the force balance: 
$\eta v_n(z=0)+F/2 \pi r \delta (z)=0$.
Here $F$ is the strength of the total force exerted by the ring of 
radius $r$ at $z=0$. 
Similar to Eq.(\ref{13}), an area integration around $z=0$ can be performed with $\rho \approx r$ and $ds=dz$. 
Then it yields the curvature as 
\be
H(z)= H_0 - \frac{F}{2\pi \kappa r}z .
\label{ring1}
\ee
where $H_0$ is the curvature at $z=0$. In this case, $H_0$ remains finite. 
The total force applied by the ring at $z=0$ is estimated from Eq.(\ref{ring1})
by taking the curvature difference as
\be
F= 2\pi r \kappa \frac{H_0-H(z_1)}{z_1}.
\label{ring2}
\ee

\begin{figure}
\begin{center} 
\includegraphics[width=4.5cm]{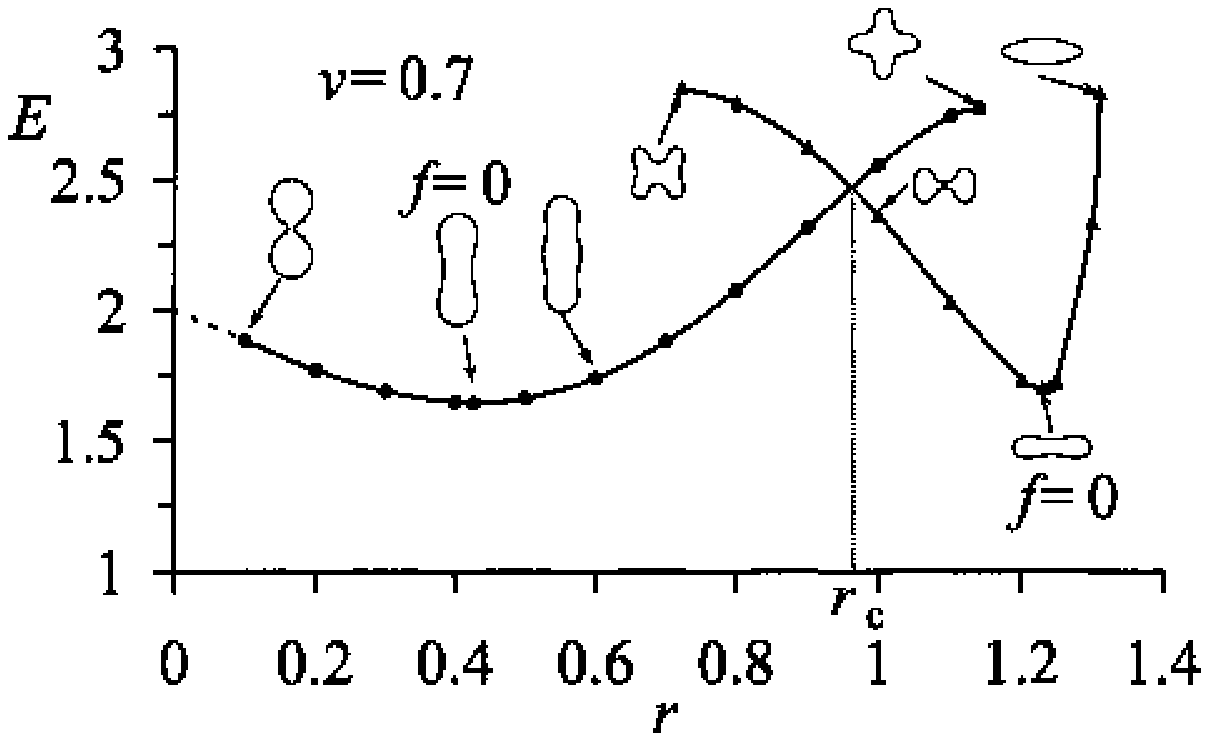}
\end{center} 
\qquad \qquad \qquad (a)
\begin{center} 
\includegraphics[width=4.5cm]{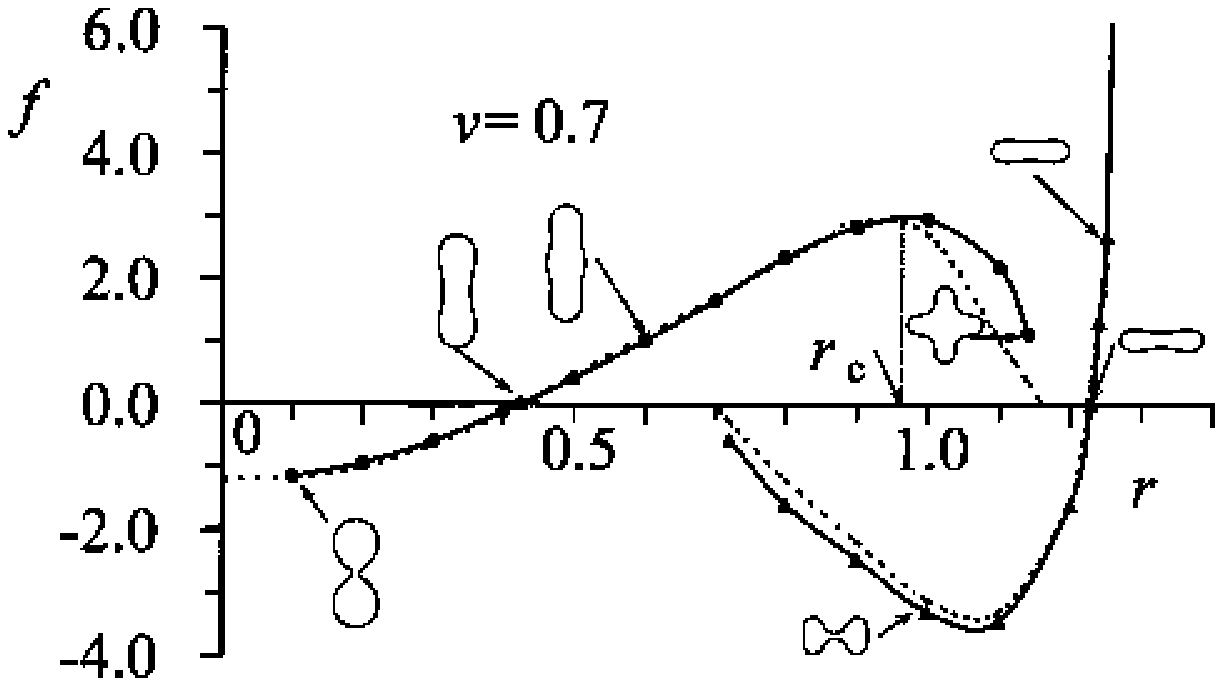}
\end{center} 
\qquad \qquad \qquad (b)
\begin{center} 
\includegraphics[width=4.5cm]{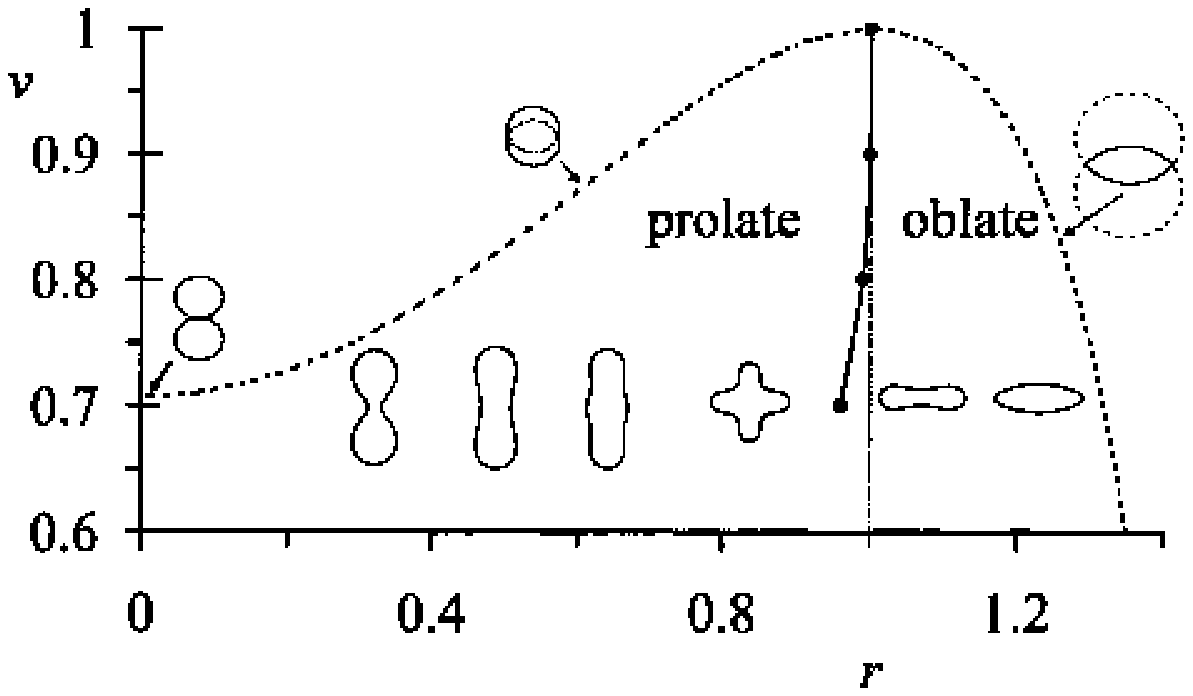}
\end{center} 
\qquad \qquad \qquad (c)
\caption{
(a) Bending energy $E$ versus ring radius $r$ 
of a vesicle with a volume $v=0.7$ 
for the prolate (filled circles) and the oblate (filled triangles) 
shapes. Lines are guides for the eyes. 
(b) The strength of the radial force $f$ versus the radius $r$ 
at the equator. A dashed lines
indicate the strength of the force induced by the derivative of 
the bending energy. 
(c) Morphological phase diagram in $v$ and $r$ space. A dashed curve 
indicateds the upper and lower limiting shapes (see text). 
The critical radius is $r_c=0.96$. 
}
\label{fig3}
\end{figure}

We now simulate the shape variation of the vesicle induced by the force 
exerted by the ring filaments. 
Every physical quantity is scaled in the dimensionless form as before.
For a given volume $v=0.7$, 
the free vesicle has a prolate shape with a 
radius about $r \approx 0.42$ at the equator $z=0$, 
and the other curvature $C_2$ is negative. 
As the ring radius increases, the bending energy increases, 
as shown in Fig.\ref{fig3}(a).
Correspondingly, the force calculated by the formula Eq.(\ref{ring2})
also increases, as shown in 
Fig.\ref{fig3}(b). 
The dotted curve shows the result obtained by numerically 
differentiating the energy shown in Fig.\ref{fig3}(a) by the radius $r$.
This indirect calculation of the force agrees well with the force
obtained directly from the simulation until the force passes the maximum for
the prolate shape and the minimum for the oblate shape.

With a small increment of the radius, 
the curvature $C_2$ at the equator is still negative. 
On increasing the ring radius, 
the central zone expands to have two positive curvatures, 
still keeping 
the two poles far apart as in the prolate shape. 
As the central radius increases, the force to expand the radius starts 
to weaken. 
At a large enough radius as $r >1.14$, the prolate becomes  unstable 
and the vesicle jumps to an oblate shape. 
The oblate at $r=1.23$ is, in fact, a shape 
with a local minimum energy without a force.
It has a bending energy being a little more than that of the 
prolate shape, and thus is a metastable shape.
 On the further increase of the ring radius, the shape 
changes to that of a flying saucer with a large energy cost. 
On decreasing the ring radius from that of the free oblate shape, 
the dips at 
poles deepen. At a small ring radius, another dips develope at the equator, 
too. Actually, the bending energies of the prolate and the oblate shapes
cross 
at a critical radius $r_c=0.96$ for the fixed volume $v=0.7$, and 
the first-order shape transition should take place there. 
The morphology phase diagram with prolate and oblate shapes is shown 
in the phase space of the volume $v$ and the ring radius $r$  as shown in 
Fig.\ref{fig3}(c). 

When the ring shrinks in the middle of the prolate shape from 
the natural radius $r=0.42$, 
as in the case of cell division, the two parts, the northern 
and the southern area, get round, as shown in Fig.3(a) and (b). 
 Since the surface area 
and the volume of the vesicle are fixed, the central radius can 
vanish only for a vesicle with a normalized volume less than 
$1/\sqrt{2} \approx 0.71$. 
When the volume is larger than this value, the shrinkage stops 
intermediately 
with a shape of two connected hemispheres. 

When the volume is smaller than this value, 
the shape of the vesicle at $r=0$ is equivalent to the two connected 
equilibrium shapes with half of the volume. 
As shown in Fig.\ref{fig3}(a), when the volume of the vesicle 
is 0.7, the vesicle approaches to the shape with
 two connected almost spherical 
shapes with $v=0.99$, where 
the bending energy of the vesicle is about 2. 

When the ring shrinks or expands in a vesicle with a fixed volume and a 
fixed surface area,
there is a certain limit for the ring radius $r$ determined solely by the
geometric reason.
For a ring with a minimum radius $r$, the vesicle takes the 
symmetrical semi-spherical shape; the radii of upper and lower spheres
are the same as $R$ and the separation between two centers is $2d$.
For a ring with a maximum radius $r$, the vesicle again takes the symmetric 
semi-spherical shape with the separation between two centers equal 
$2d$ being negative.
The area conservation gives the condition $4 \pi R(R+d)=4 \pi$, and the 
volume conservation $2 \pi(R+d)^2(2 R- d)/3=4 \pi v/3$.
The  radius at the equator is given by $r=\sqrt{2- R^{-2}}$.
When $d=R=1/\sqrt{2}$, the vesicle consists of two complete spheres with $r=0$
and the volume is equal $v=1/\sqrt{2}$.
When $d=-R=-\infty$, the vesicle becomes flat with a vanishing volume $v=0$,
 but the radius at the equator remains finite $r=\sqrt{2}$. 
 The upper and lower limits of $r$ for arbitrary volume $v$ can be easily 
 obtained numerically, and are shown by a dashed
curve in Fig.\ref{fig3}(c). 
At $v=0.7$, the upper limit of the radius $r$ thus obtained agrees well
with the result of dynamical simulation.

\section{Impurity Diffusion}
\label{sec5}

There are usually various ions or proteins dispersed in a membrane, 
and they affect the shape of vesicles. 
When the impurities are homogeneously distributed, the vesicle 
takes a shape determined by the curvature elasticity with some 
modification in the spontaneous curvature. But when the impurities 
are coupled to the local curvature of the membrane, the homogeneous 
distribution may become unstable. The phase separation 
and the shape instability are induced 
simultaneously. We here study this phenomenon 
analytically and numerically. 

\subsection{Dynamics driven by impurities}
Since the homogeneously distributed impurities only modify the spontaneous 
curvature and the surface tension $\sigma$, we set the average concentration 
to be zero.
The local concentration fluctuation of intercalated 
molecules from the average 0
is denoted by $\Phi(\vec r,t)$. 
The interaction of intercalated molecules with the phospholipidic 
constituents of the membrane might be summarized in the form,
\be
F_{int} = - \Lambda \int d^D s \sqrt{g} H \Phi .
\label{16}
\ee
with a coupling constant $\Lambda$. 
The free energy of the impurity molecules is assumed to be
Landau-Ginzburg form as
\be
F_{imp} = \int d^Ds \sqrt{g} [ \frac{1}{2} B 
(\vec \nabla \Phi)^2+ f(\Phi)] ,
\label{17}
\ee
where the homogeneous part is expanded up to the fourth order as 
\be
f(\Phi)= \frac{A_2}{2} \Phi^2 + \frac{A_4}{4} \Phi^4 .
\label{18}
\ee
Since the average fluctuation of impurity concentration is set to zero, 
there is no linear term of $\Phi$ in the free energy $f(\Phi)$.
The evolution dynamics of the membrane, 
Eq.(\ref{3}) now contains terms due to the impurity as 
\bea
v_n &= v_c 
+ \eta^{-1} \Lambda \{ 2C_1C_2 \Phi - \tilde \Delta \Phi \}
- \eta^{-1} \{ \frac{B}{2} (\partial_s \Phi)^2 (C_1-C_2)
\nonumber \\
& + ( \frac{A_2}{2} \Phi^2 + \frac{A_4}{4} \Phi^4 ) H \}
-\eta^{-1} \sigma H -\eta^{-1} P .
\label{19}
\eea
The conservation of the total number of 
impurity molecules on the membrane 
leads to the restriction $\int d^Ds \sqrt{g} \Phi =0$. 
Therefore, the time-evolution of the impurity 
concentration fluctuation $\Phi$ should satisfy 
\be
0= \frac{d}{dt} \int d^Ds \sqrt{g} \Phi = \int d^Ds \sqrt{g}
( \frac{\partial \Phi}{\partial t} 
 + \Phi H v_n)  .
\label{20}
\ee
This is satisfied by describing the evolution in terms of the 
diffusion equation as 
\be
\partial _t \Phi + \Phi H v_n = 
D \tilde \Delta \left( \frac{1}{\sqrt{g}} 
\frac{ \delta F}{\delta \Phi}
\right) .
\label{21}
\ee
In a simulation where a membrane is discretized into grid points, 
the supporting grid points have a tangential velocity in addition to 
the physical normal velocity, so as to keep the grid separation 
equidistant. Accordingly, the concentration fluctuation field $\Phi$ should 
include this contribution given as
\be
\delta \Phi = \frac{\partial \Phi (\vec{r}(t))}{\partial t} 
\ \delta t = \frac{\partial \vec{r}}{\partial t} 
\cdot \nabla \Phi \ \delta t = v_t \partial _s \Phi \ \delta t .
\ee
Therefore, the  final evolution for $\Phi$ is described as 
\be
\partial _t \Phi + \Phi H v_n - 
v_t \partial _s \Phi = 
 D \tilde \Delta (-B \tilde \Delta \Phi + A_2 \Phi + A_4 \Phi^3 
- \Lambda H) .
\label{22}
\ee

We introduce the following dimensionless parameters hereafter.
\bea
&&\phi=\Phi R_0^2, \quad
b=\frac{B}{\kappa R_0^4},\quad a_2=\frac{A_2}{\kappa R_0^2},\quad a_4
=\frac{A_4}{\kappa R_0^6},
\nonumber \\
&& \quad \lambda 
=\frac{\Lambda}{\kappa R_0},\quad d=\eta D R_0^4,\quad c_0=C_0 R_0 .
\label{23}
\eea
\subsection{Linear stability analysis}
In this section we study the stability of the spherical vesicle
with homogeneously distributed impurities under a small shape fluctuation
and the impurity inhomogeneity. 
For simplicity, we consider the case with a fixed pressure $P$, and thus 
the volume of the vesicle can be altered. 
In the unperturbed situation, the vesicle has a radius $R_0$.
Assuming the small deformation of the vesicle and no overhang configuration,
the vesicular position vector $\vec r$  and 
the concentration of impurities $\phi$ in dimensionless form are 
single-valued functions of the polar  $\theta$ and the azimuthal angle 
$\varphi$  in the spherical coordinate. 
These small deviations are expanded in terms 
of spherical harmonic function  $Y^l_m$, and we note that 
each mode is independent in the linear stability analysis.
Therefore, one considers only the $l$-pole mode, and 
the membrane position vector $\vec r$ and 
the impurity concentration $\phi$ in dimensionless form
are written as 
\bea
\vec{r}(\theta ,\varphi ,t) &= \{ 1+\delta ^l_m(t)
Y^l_m(\theta ,\varphi )\} \vec{e}_r(\theta ,\varphi ) , 
\\
\phi (\theta ,\varphi ,t) &= \zeta ^l_m(t)Y^l_m(\theta ,\varphi ) .
\label{24}
\eea
where $\delta$ and $\zeta$ describe the time-dependent coefficients 
of fluctuations and $\vec e_r$ is the unit radial vector.
Since the average concentration is set to zero, there is no unperturbed term
in the expansion of the concentration $\phi$.
From Eqs.(\ref{19}) and (\ref{22}), one obtains 
the linear dynamics of 
the perturbation $\delta ^l_m$, $\zeta ^l_m$ as 
\begin{widetext}
\bea
\label{24a}
\frac{\partial \delta ^l_m}{\partial t} &\simeq &
 (l+2)(l-1)\left\{ -l(l+1)+\frac{1}{2}p+c_0 
\right\} \delta ^l_m 
+ \left\{ 2\lambda +\lambda l(l+1) \right\} \zeta ^l_m ,
 \\ 
\frac{\partial \zeta ^l_m}{\partial t} &\simeq &
 d \left\{ -l(l+1) \left[ b l(l+1)+a_2  \right] \zeta ^l_m 
+\lambda l(l+1)(l+2)(l-1)\delta ^l_m \right\} .
\label{24b}
\eea
The conservation of the surface area determines the surface tension $\sigma$.
With a small shape deformation $\delta$, the variation of $\sigma$ is easily
shown to be quadratic in $\delta$, and it does not affect the linear stability
analysis.
Eqs. (\ref{24a}) and (\ref{24b}) can be reduced to matrix from as follows. 
\bea
\left(
\begin{array}{c}
\partial _t\delta ^l_m \\
\partial _t\zeta ^l_m \\
\end{array}
\right) 
&=
\left(
\begin{array}{cc}
-(l+2)(l-1) \left\{ l(l+1)-\frac{p}{2}-c_0 \right\} 
& \{ 2+l(l+1)\} \lambda  \\
dl(l+1)(l+2)(l-1)\lambda & -dl(l+1)\{ bl(l+1)+a_2 \} 
\end{array}
\right) 
\left(
\begin{array}{c}
\delta ^l_m \\
\zeta ^l_m \\
\end{array}
\right) \nonumber \\
&\equiv 
\left(
\begin{array}{cc}
M_{rr} & M_{r\phi } \\
M_{\phi r} & M_{\phi \phi } 
\end{array}
\right) 
\left(
\begin{array}{c}
\delta ^l_m \\
\zeta ^l_m \\
\end{array}
\right) .
\label{25}
\eea
Note that the time-development matrix $M$ 
is independent of $m$.
The eigenvalues $\beta _{+}, \beta _{-}$ and eigenvectors 
$(\delta _{\pm}, \zeta _{\pm})$ of the matrix $M$ define 
the time evolution of the deviations $\delta ^l_m, \zeta ^l_m $ 
as follows. 
\bea
\left(
\begin{array}{c}
\delta ^l_m \\
\zeta ^l_m \\
\end{array}
\right) 
&=
A_{+}\exp (\beta _{+}t)
\left(
\begin{array}{c}
\delta _{+} \\
\zeta _{+} \\
\end{array}
\right) 
+
A_{-}\exp (\beta _{-}t)
\left(
\begin{array}{c}
\delta _{-} \\
\zeta _{-} \\
\end{array}
\right) ,
\\
&\beta _{\pm}=\frac{1}{2}\{ M_{rr}+M_{\phi \phi }\pm 
\sqrt{(M_{rr}-M_{\phi \phi })^2+4M_{r\phi }M_{\phi r}} \}
, 
\nonumber \\
&\left(
\begin{array}{c}
\delta _{\pm} \\
\zeta _{\pm} \\
\end{array}
\right) 
=
\left(
\begin{array}{c}
M_{r\phi } \\
-M_{rr}+\beta_{\pm} \nonumber \\
\end{array}
\right) .
\label{26}
\eea
\end{widetext}

%

\begin{figure}
\begin{center}
\includegraphics[width=8.5cm]{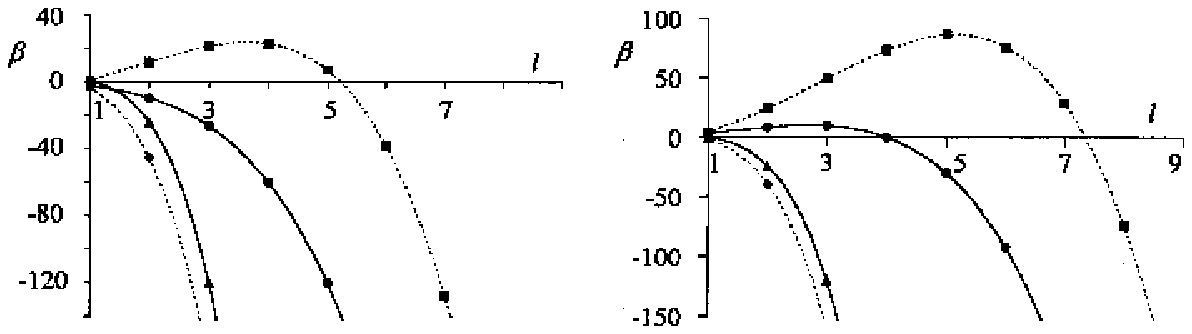}%
\\
(a) \qquad \qquad \qquad 
\qquad \qquad (b)
\end{center} 
\caption{
Growth rate $\beta$ as a function of mode number $l$ in the linear 
stability analysis for P=0, $C_0=0$, 
(a) for $a_2>0$, (b) for $a_2<0$. 
Solid curves represent the 
growth rates for $\Lambda = 0$, 
$\beta_{+}$ (filled circles) and $\beta_{-}$ (filled triangles).
Dashed curves represent the growth rates for $\Lambda \ne 0$, 
$\beta_{+}$ (filled squares) and $\beta_{-}$ (filled diamonds).
The parameters are $b = 0.1,\  a_2 = 1 \ $or$\ -2,\  a_4 = 1,\  
\lambda =2.0\  d = 1,\  c_0 = 0, p=0$. 
}
\label{fig4}
\end{figure}

In Fig.\ref{fig4} we show the relationship between the growth rate 
$\beta _{+}, \beta _{-}$ and mode number $l$ for both cases 
without $\Lambda = 0$ (solid curves)  and with the 
coupling $\Lambda \neq 0$ (dashed curves). 
\par

With a negative $a_2$, as in Fig.\ref{fig4}(b) 
impurities can phase separate spontaneously, 
i.e. even if there is no coupling to 
the curvature of the membrane, $M_{\phi \phi} >0$ for $l^{\ast} \ge 1$
when $2b < -a_2$. 
Then the most unstable mode $l^{\ast}_{\Lambda = 0}$ 
is the integer close to 
$1 / 2(-1 + \sqrt{1 - 2a_2/ b})$. 
When $2b < -a_2 \le 6b$, the most unstable mode is $l^{\ast}=1$. The 
impurities tend to phase separate into two domains. With a small 
coupling to the membrane curvature ($\Lambda \ne 0$), 
this mode induces the translation of the vesicle  without deformation. 
When $6b < -a_2$, some modes with $l^{\ast} \ge 2$ becomes unstable. 
With a small coupling $\Lambda$, the vesicle  deforms accordingly. 

To confirm the linear stability analysis and further to elucidate 
the nonlinear effect, we simulate the dynamical evolution of the 
vesicular shape and the impurity diffusion. 
We assume axisymmetry in the simulation, and thus only $m=0$ modes are 
relevant here.
The parameters chosen are  
$b = 0.1,\  a_2 = -1,\  a_4 = 1,\  \lambda =3.0, 
\  d = 1,\  c_0 = 0$. 
The pressure $p$ is fixed to be 
$p = 0$. 
The most unstable mode is expected to be $l^{\ast}=6$. 

\begin{figure}
\begin{center} 
\includegraphics[width=8.5cm]{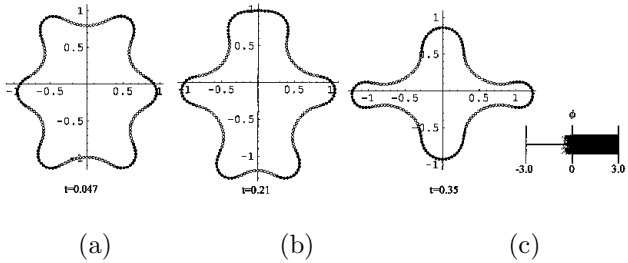}
\\
(a) \qquad \qquad \qquad (b) \qquad \qquad \quad \qquad (c)
\end{center} 
\caption{
Time evolution of the vesicle with impurities, 
the parameters are $b = 0.1,\  a_2 = -1,\  a_4 = 1,\  
\lambda =3.0\  d = 1,\  c_0 = 0, p=0$. 
The most unstable mode is $l^{\ast}=6$ 
The time is measured in unit of $\eta R_0^4/\kappa $. 
}
\label{fig5}
\end{figure}

The initial shape is spherical, and the 
impurities are distributed randomly whose Fourier amplitudes up 
to 32nd mode $\zeta_0^{\ell}$ for $\ell \le 32$
have uniformly  distributed 
between $-0.001$ and $0.001$. 
In the course of simulation the vesicle shape 
developes the wavy pattern with 6 domains of impurities,
as shown in Fig.\ref{fig5}(a). 
Impurities accumulate to the convex portion of the vesicle due to
the positive coupling $\Lambda>0$.
This number corresponds to the most 
unstable mode with $l^{\ast} =6$. 
As the time evolves, these domains coalesce each other. 
In last stage of Fig.5(c) the number of the domains becomes 4.
%

\begin{figure}
\begin{center} 
\includegraphics[width=8.5cm]{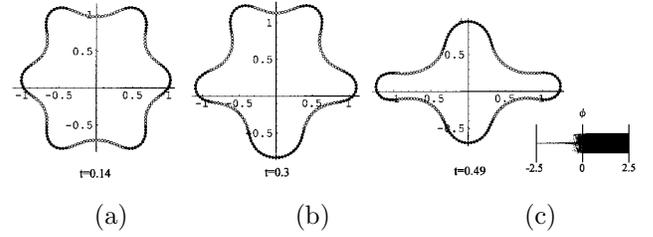}%
\\
(a) \qquad \qquad \qquad (b) \qquad \qquad \quad \qquad (c)
\end{center} 
\caption{
Time evolution of the vesicle with impurities, 
the parameters are $b = 0.1,\  a_2 = 1,\  a_4 = 1,\  
\lambda =3.0\  d = 1,\  c_0 = 0, p=0$. 
The most unstable mode is $l^{\ast}=6$ 
}
\label{fig6}
\end{figure}

\begin{figure}
\begin{center} 
\includegraphics[width=8.5cm]{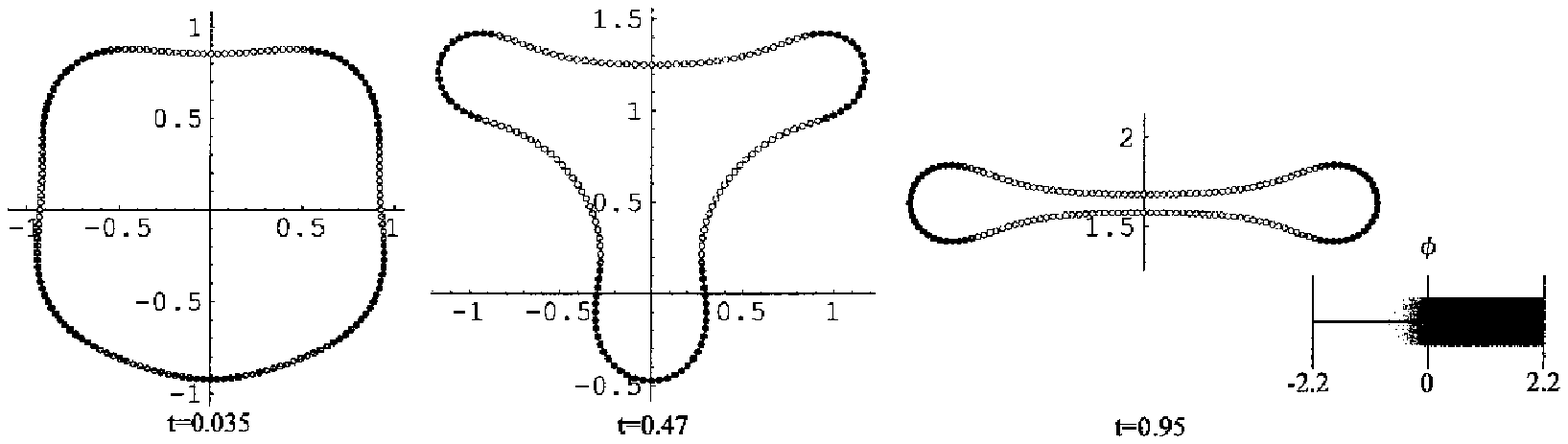}%
\\
(a) \qquad \qquad \qquad (b) \qquad \qquad \quad \qquad (c)
\end{center} 
\caption{
Time evolution of the vesicle with impurities, 
the parameters are $b = 0.1,\  a_2 = 1,\  a_4 = 1,\  
\lambda =2.5\  d = 1,\  c_0 = 0, p=0$. 
The most unstable mode is $l^{\ast}=5$ 
}
\label{fig7}
\end{figure}

\begin{figure}
\begin{center} 
\includegraphics[width=8.5cm]{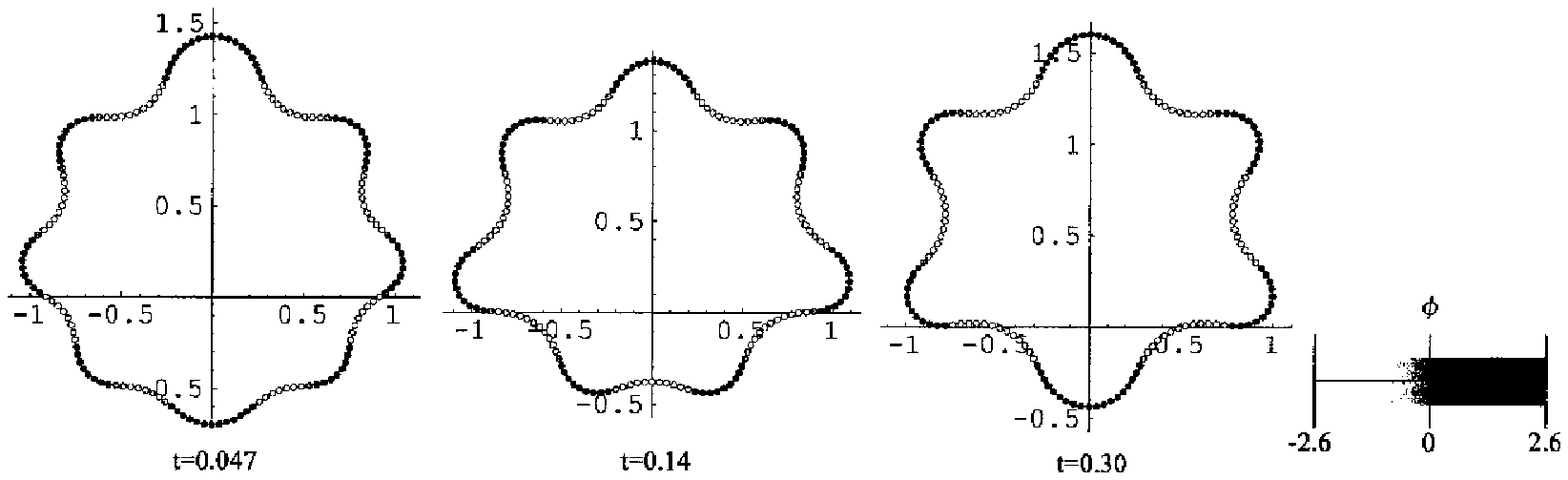}%
\\
(a) \qquad \qquad \qquad (b) \qquad \qquad \quad \qquad (c)
\end{center} 
\caption{
Time evolution of the vesicle with impurities, 
the parameters are $b = 0.05,\  a_2 = 1,\  a_4 = 1,\  
\lambda =3.0\  d = 1,\  c_0 = 0, p=0$. 
The most unstable mode is $l^{\ast}=9$ 
}
\label{fig8}
\end{figure}

\begin{figure}
\begin{center} 
\includegraphics[width=8.5cm]{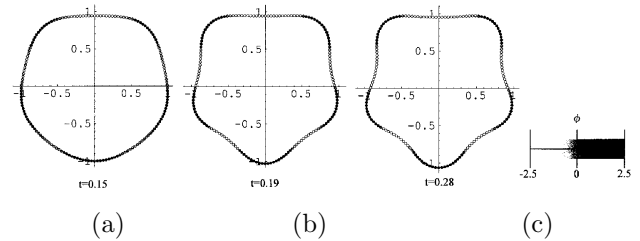}%
\\
(a) \qquad \qquad \qquad (b) \qquad \qquad \quad \qquad (c)
\end{center} 
\caption{
Time evolution of the vesicle with impurities, 
the parameters are $b = 0.1,\  a_2 = 1,\  a_4 = 1,\  
\lambda =3.0\  d = 1,\  c_0 = 0, p=-40$. 
The most unstable mode is $l^{\ast}=5$ 
}
\label{fig9}
\end{figure}

With a positive $a_2$, accumulation of impurities costs energy and 
they try to diffuse away to be homogeneous if there is no coupling 
to the curvature of the membrane. 
However, due to the coupling between the concentration variation and
the membrane curvature, the homogeneous distribution can be unstable
and the phase separation is possible for sufficiently strong coupling 
as shown in Fig.\ref{fig4}(a). 
The linear stability analysis shows that 
a spherical vesicle becomes unstable if the coupling is 
stronger than the critical value 
$ |\lambda ^{\ast}| = \sqrt{18b + 3a_2} / 2$. 
For the parameter choice 
$b = 0.1,\ a_2 = 1,\  a_4 = 1,\  d = 1,\  c_0 = 0$, 
the critical coupling takes the value $\lambda  ^{\ast}\sim 1.09$. 
When the coupling is stronger than $\lambda ^{\ast}$ such as 
$\lambda =3.0$, the most unstable mode has 
$l^{\ast} = 6$, in good agreement to the early stage of the simulation result 
as shown in Fig.\ref{fig6}(a). 

By decreasing  the coupling such as $\lambda =2.5$, 
the number of the most unstable mode decreases to
$l^{\ast} = 5$, in agreement to the simulation result
shown in Fig.\ref{fig7}(a). 
When the correlation length $b$ gets smaller such as $b=0.05$, 
the number of the most unstable mode increases as 
$l^{\ast} = 9$. In the simulation the vesicle becomes 
wavy with 8 domains in the early stage 
as shown in Fig.\ref{fig8}(a). 
The result has the same tendency with the linear stability analysis. 
In all the cases studied, the coarsening takes place as the time elapses,
and the number of domains decreases.

We can study the effects of the pressure $p$ or 
the spontaneous curvature $c_0$  by linear analysis. 
According to Eq.(\ref{25}), the positive $p$ or $c_0$ increase 
the number of the most unstable mode $l^{\ast}$. 
On the contrary the negative $p$ or $c_0$ decrease $l^{\ast}$. 
For example, we simulated 
the shape of the vesicle with the parameters such as 
$p=-40,\  b = 0.1,\  a_2 = 1,\  a_4 = 1,\  
\lambda =3.0\  d = 1,\  c_0 = 0$ 
as shown in Fig.\ref{fig9}. Then  
the linear analysis gives the most unstable mode $l^{\ast}=5$,
in fair agreement with the simulation. 
From Fig.9(b) and (c), the coarsening does not seem to
take place under the pressure. 
This might be related to the resistence to the volume change at 
a negative $p$.

In almost all cases the results of the simulation in the early stage
agree with the linear stability analysis. 
Moreover, the phase separation domains coalesce each other, 
as the time evolves. 
This nonlinear coalescing process is also observed in the previous 
study\cite{Kumar+01}. 

\section{Conclusion}
\label{sec6}

We performed simulations of the axisymmetric vesicle shape change caused by the
length change of an encapsulated  ring or rod filaments. 
The constant volume simulation of the vesicle with a rod encapsulation
shows that the $\phi$ shape consisting of a spherical and a tubular
portions is attributed to the restriction of the volume and the surface area
conservations.
The same restriction limits the extention of the ring radius in the case
of the ring-filament encapsulation.
The shape transitions of the vesicle are summarized 
in phase diagrams, Fig.2(b) and 3(c), 
in the phase space of the vesicular volume 
and a rod-filament length or a ring-filament radius. 
\cite{umeda+98,morikawa+99,fygenson+99,heinrich+99}. 

We also performed simulations on the dynamics of the vesicle shape 
caused by the
phase separation of the impurity concentration.
Impurities which show spontaneous phase separation induces periodic 
deformation in the vesicle shape. 
Even those impurities which do not give rise to spontaneous 
phase separation are shown 
to induce periodic deformations through the strong coupling of the impurity
concentration to the membrane curvature.
The early stage is well explained by the linear stability analysis, such
as the dominant periodicity of the shape deformation.
In late stage where the nonlinear effect becomes dominant,
the coarsening of the shape periodicity takes place.
Similar coasening process was observed in the previous 
study\cite{Kumar+01}. 
The study on the strongly deformed vesicle due to the impurity
phase separation is now under way.


\begin{thebibliography}{99}
\bibitem{helfrich73}
W. Helfrich: Z. Naturforsch. C{\bf 28}, (1973) 693.
\bibitem{udo+91}
U. Seifert, K. Berndl, and R. Lipowsky: Phys. Rev. A{\bf 44}, (1991) 1182.
\bibitem{bray92}
D. Bray: {\it Cell Movements} (Garland Publishing, New York, 1992).
\bibitem{hotani+90}
H. Hotani, and H. Miyamoto: Adv. Biophys. {\bf 26}, (1990) 135.
\bibitem{umeda+98}
T. Umeda, H. Nakajima, and H. Hotani: J. Phys. Soc. Jpn. {\bf 67}, (1998) 682.
\bibitem{morikawa+99}
R. Morikawa, Y. Saito, and H. Hyuga: J. Phys. Soc. Jpn. {\bf 68}, (1999) 1760.
\bibitem{fygenson+99}
D. K. Fygenson, J. F. Marko, and A. Libehaber: Phys. Rev. Lett. {\bf 79}, (1997) 4497.
\bibitem{heinrich+99}
V. Heinrich, B. Bozic, S. Svetina and B. Zeks: Biophys. J. {\bf 76}, (1999) 2056. 
\bibitem{miyata+92}
H. Miyata, and H. Hotani: Proc. Natl. Acad. Sci. USA {\bf 89}, (1992) 11547.
\bibitem{deuticke+68}
B. Deuticke: Biochem. Biophys. Acta. {\bf163 }, (1968) 494.
\bibitem{allan+75}
D. Allan, and R. H. Michell: Nature {\bf 258}, (1975) 348.
\bibitem{gebhardt+77}
C. Gebhardt, H. Gruler, and E. Sackmann: Z. Naturforsch. {\bf32c}, (1977) 581.
\bibitem{bradley+99}
A. J. Bradley, E. Maurer-Spurej, D. E. Brooks, and D. V. Devine: Biochemistry {\bf 38}, (1999) 8112.
\bibitem{tsafrir+01}
I. Tsafrir, D. Sagi, T. Arzi, M-A. G-Boudeville, V. Frette, D. Kandel, and J. Stavans: Phys. Rev. Lett. {\bf86}, (2001) 1138.
\bibitem{leibler86}
S. Leibler: J. Physique {\bf 47}, (1986) 507.
\bibitem{taniguchi+96}
T. Taniguchi: Phys. Rev. Lett. {\bf 76}, (1996) 4444.
\bibitem{Kumar+01}
P.B. S. Kumar, G. Gompper, R. Lipowsky: Phys. Rev. Lett. {\bf 86}, (2001) 3911.
\bibitem{marsili+96}
M. Marsili, A. Maritan, F. Toigo, and J. R. Banavar: Rev. Mod. Phys. 
{\bf 68}, (1996) 963.
\bibitem{langer+92}
S. Langer, R. Goldstein and D.P. Jackson: Phys. Rev. A{\bf 46}, (1992) 4894.
\bibitem{cantat+00}
I. Cantat, C. Misbah, and Y. Saito: Eur. Phys. J. E{\bf 3}, (2000) 403.


\end{thebibliography}
\end{document}